# MusicRWKV: A Multimodal RWKV Framework for Music Representation Learning

## Abstract

Recent music foundation models have achieved remarkable progress in music representation learning, yet most existing approaches rely on Transformer architectures whose quadratic attention complexity introduces substantial computational and memory overhead when modeling long-duration music recordings. In this work, we propose MusicRWKV, a multimodal music representation learning framework built upon the RWKV architecture. MusicRWKV jointly models waveform, spectrogram, and metadata information within a unified self-supervised learning framework. To improve efficiency for long musical sequences, we introduce an invertible compression module that reduces sequence length while preserving reconstruction capability. A FiLM-based conditioning strategy is employed to facilitate modality-aware representation learning without introducing expensive cross-attention operations. We further investigate the integration of Soft ROSA as a differentiable retrieval mechanism for neural-symbolic interaction. Experimental evaluation across a diverse collection of Music Information Retrieval (MIR) benchmarks demonstrates that the proposed RWKV-based architecture can be effectively adapted to heterogeneous music understanding tasks. A controlled pretraining ablation shows that Soft ROSA improves validation performance, while layer-wise telemetry analysis reveals a progressive transition from distributed retrieval to selective retrieval in deeper network layers. These results suggest that RWKV provides a promising architectural alternative to Transformer-based models for multimodal music understanding.

# 1 Introduction

Learning music representation has become a fundamental problem in Music Information Retrieval (MIR). Recent advances in self-supervised learning (SSL) have enabled the development of large-scale music foundation models that learn transferable representations from unlabeled audio data. Models such as MERT [1], MusicFM [2], and MuQ [3] have demonstrated strong generalization ability across a diverse range of downstream MIR tasks, including music tagging, genre classification, key estimation, emotion recognition, and instrument classification. These models are typically pretrained on large collections of music recordings and later evaluated through linear probing or lightweight task-specific adaptation.

Despite their success, existing music representation learning frameworks, including MERT, MusicFM, and MuQ, are predominantly based on Transformer architectures. While Transformers provide powerful long-range sequence modeling capabilities, their quadratic attention complexity introduces substantial computational overhead when modeling long-duration music recordings. In addition, Transformer inference commonly relies on Key-Value (KV) caching, whose memory consumption grows with sequence length. For music recordings that may span several minutes, high-resolution audio tokenizers may produce extremely long token sequences. In such scenarios, the accumulation of KV caches can become a substantial memory bottleneck during both training and deployment.

Recently, a new generation of efficient foundation model architectures has emerged beyond the conventional Transformer paradigm. Among them, state-space and recurrent architectures have attracted increasing attention due to their ability to model long-context sequences with substantially lower computational complexity. The RWKV architecture [4] is one representative example of this trend, combining the parallelizable training characteristics of Transformers with the linear-time inference complexity of recurrent neural networks. Recent generations of RWKV have demonstrated competitive performance against Transformer-based large language models while offering superior memory efficiency and scalability for long-context modeling. These advantages make RWKV particularly appealing for music understanding, where audio recordings often contain long temporal dependencies and rich hierarchical structures spanning rhythm, timbre, harmony, and semantics.

Furthermore, recent work has explored RWKV for symbolic music generation through the MIDI-RWKV [5] framework, demonstrating the effectiveness of RWKV for long-context symbolic music modeling. However, existing RWKV-based music studies mainly focus on symbolic sequence generation and controllable music creation rather than learning transferable audio representations for downstream music understanding tasks. To the best of our knowledge, the potential of RWKV as a backbone architecture for self-supervised music representation learning remains largely unexplored.

Another challenge in music representation learning lies in the heterogeneous nature of musical information. Music understanding often requires integrating information from multiple modalities, including raw audio waveforms, spectral representations, and semantic metadata. Existing Transformer-based music foundation models frequently employ cross-attention mechanisms to facilitate modality interaction. In contrast, RWKV does not inherently provide a cross-attention framework for multimodal fusion, making effective integration of heterogeneous musical representations an open problem.

To address these limitations, we propose **MusicRWKV**[1], a multimodal RWKV framework for music representation learning. The proposed framework jointly models waveform tokens, spectrogram tokens, and metadata tokens within a unified RWKV architecture. To efficiently handle long musical sequences, we introduce an invertible compression module that reduces sequence length while preserving reconstruction capability. A FiLM-based [6] conditioning strategy is adopted to provide modality-aware representation learning without introducing expensive cross-attention operations. Furthermore, we investigate the effect of differentiable neural-symbolic interaction by incorporating the Soft ROSA [7] mechanism into RWKV blocks as an optional enhancement for representation learning.

To evaluate the applicability of the proposed architecture across diverse MIR scenarios, we conduct downstream experiments on music tagging, genre classification, key estimation,

[1] code repository: https://github.com/tanjiarui/music-rwkv

emotion recognition, instrument classification, and singing voice analysis. For each task, we train a task-specific MusicRWKV model from scratch using a shared backbone design, while adapting the input modalities and output heads to the corresponding task.

The main contributions of this work are summarized as follows:

1. To the best of our knowledge, MusicRWKV is the first multimodal RWKV framework specifically designed for audio-based music representation learning.
2. We demonstrate that RWKV can serve as an effective alternative to Transformer-based architectures for self-supervised music representation learning while retaining efficient long-sequence modeling capability.
3. We construct a unified multimodal pretraining framework that jointly models waveform, spectrogram, and metadata information for music understanding.
4. We provide an empirical analysis of Soft ROSA through ablation experiments and layer-wise telemetry statistics.

# 2 Related Work

## 2.1 Music Representation Learning

Music representation learning aims to learn transferable semantic representations from large-scale unlabeled music data and has become a fundamental research direction in Music Information Retrieval (MIR). Early studies commonly adopted self-supervised learning paradigms based on contrastive learning. Representative approaches such as CLMR [8] and MULE [9] demonstrated the effectiveness of contrastive objectives for learning reusable music embeddings and established the foundation of self-supervised music representation learning.

With the development of foundation models, recent research has increasingly focused on large-scale self-supervised pretraining. Among recent music foundation models, MERT introduced a masked prediction framework combining acoustic and musical teachers to learn transferable music representations. MusicFM further explored a unified foundation model for music informatics tasks, while MuQ adopted Mel Residual Vector Quantization (Mel-RVQ) as a self-supervised target and achieved state-of-the-art performance across multiple MIR benchmarks. More recently, multitask self-supervised learning frameworks such as MT2 [10] demonstrated that combining multiple pretext objectives can further improve representation quality and transferability across heterogeneous music understanding tasks.

Despite their differences in architecture design and training objectives, most contemporary music representation learning frameworks remain based on Transformer architecture. Consequently, the potential of alternative foundation model architectures for music representation learning remains insufficiently explored.

## 2.2 Multimodal Music Understanding

Music understanding is inherently multimodal. Beyond raw audio signals, musical information can be represented through multiple complementary modalities, including waveform representations, spectrogram representations, and semantic metadata. Each modality captures different aspects of musical content. Waveforms preserve fine-grained temporal information, spectrograms emphasize frequency-domain structures such as timbre and harmony, while metadata provides high-level semantic descriptions regarding genre, mood, instrumentation, and contextual information.

Recent studies have demonstrated that integrating multiple modalities can improve the quality of learned music representations. In addition, several works have explored structured representation learning from multiple views. For example, MARL [11] investigates disentangled music representations by separating shared and view-specific information across different audio views. Such studies highlight the importance of exploiting heterogeneous musical information for robust representation learning. However, existing multimodal music learning frameworks are predominantly designed around attention-based architectures, leaving multimodal learning in RWKV-based systems largely unexplored.

## 2.3 Efficient Foundation Model Architectures

Transformer architecture has become the dominant paradigm for foundation models due to their strong sequence modeling capability. However, the quadratic computational complexity of self-attention introduces substantial memory and computational overhead when processing long-context sequences. This limitation becomes increasingly problematic in music understanding scenarios, where audio recordings often span several minutes and contain rich hierarchical structures.

To address this challenge, recent studies have explored more efficient foundation model architectures, including state-space models and recurrent architectures. These approaches aim to retain long-range dependency modeling capabilities while reducing computational complexity. Among them, RWKV has emerged as a promising foundation model architecture that combines Transformer-style parallel training with recurrent inference. By maintaining linear complexity with respect to sequence length during inference, RWKV provides a scalable solution for long-context sequence modeling while achieving competitive performance against Transformer-based architectures.

## 2.4 RWKV for Music Modeling

Although RWKV has attracted considerable attention in language modeling, its application to music remains relatively limited. Existing music-related RWKV research primarily focuses on symbolic music modeling. The most representative example is Adaptable Symbolic Music Infilling with MIDI-RWKV, which employs the RWKV-7 [12] architecture for long-context symbolic music infilling and controllable music generation. Experimental results demonstrate that RWKV can effectively model long-range symbolic musical structures and generate coherent musical sequences.

However, symbolic music generation and audio representation learning address fundamentally different objectives. Symbolic generation focuses on autoregressive modeling of discrete musical events, whereas music representation learning aims to learn transferable features that generalize across diverse downstream MIR tasks. To the best of our knowledge, existing RWKV-based music studies have not systematically investigated RWKV as a backbone architecture for multimodal self-supervised music representation learning. In contrast, our work explores RWKV for audio-based multimodal music representation learning and evaluates the applicability of the proposed backbone architecture across a broad collection of MIR benchmarks.

# 3 Methodology

Before describing each component in detail, we first provide an overview of the data flow in MusicRWKV. Waveform, spectrogram, and metadata tokens are processed by modality-specific branches, fused into a shared representation space, and optimized through self-supervised objectives. The following subsections describe each component in detail.

## 3.1 Overview

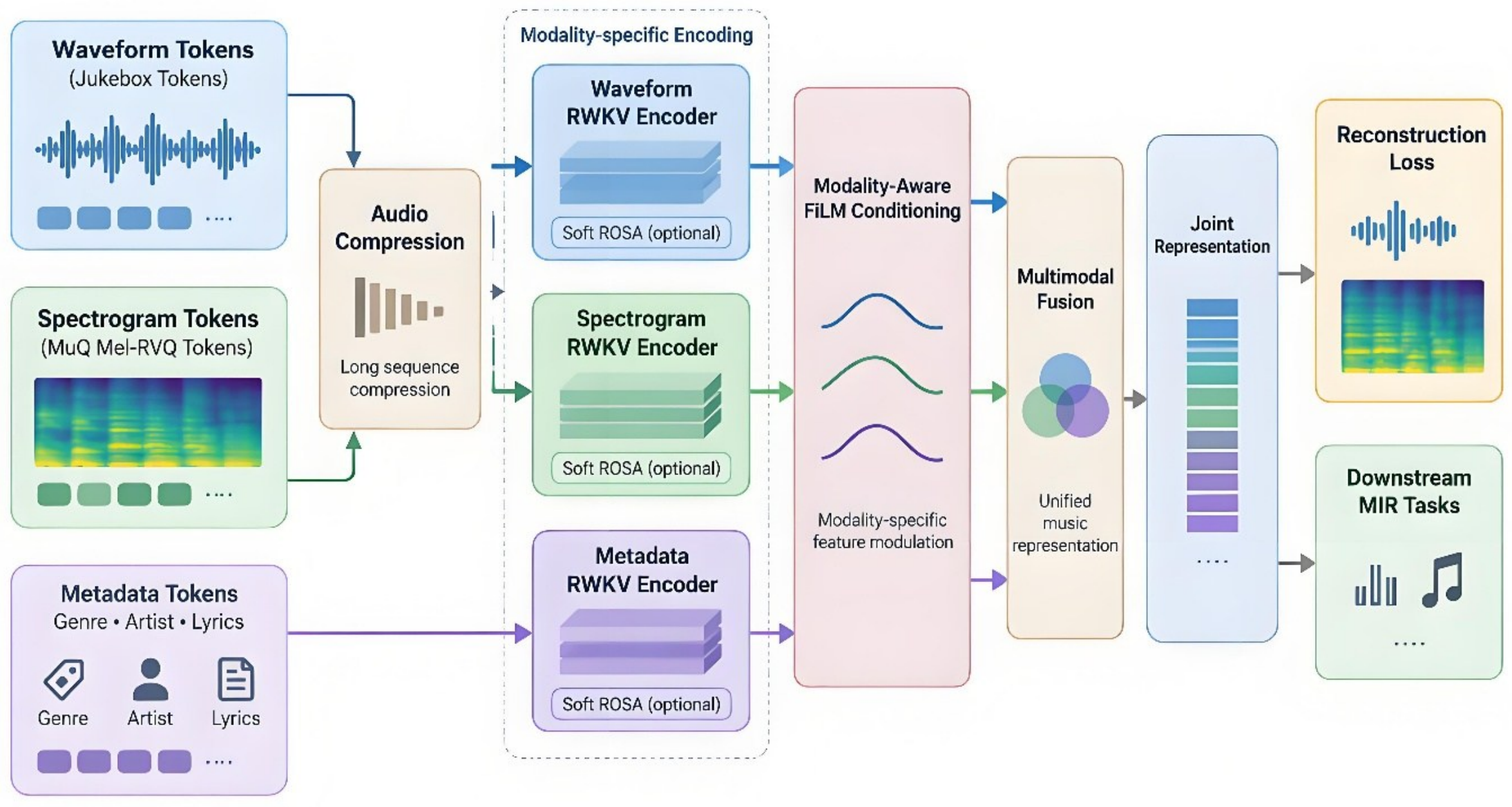

Figure 1: Illustration of MusicRWKV Framework

Figure 1 illustrates the overall architecture of MusicRWKV, a multimodal framework for music representation learning built upon the RWKV architecture. The framework jointly models three complementary modalities of musical information, including waveform tokens, spectrogram tokens, and semantic metadata.

To efficiently process long-duration music recordings, waveform and spectrogram token sequences are first passed through an invertible audio compression module that reduces sequence length while preserving reconstruction capability. This design reduces the effective sequence length presented to the RWKV encoders, which is particularly important for long-duration music recordings. Metadata tokens are directly forwarded to the representation encoder without compression. The resulting modality-specific token sequences are then processed by independent RWKV encoders to capture temporal and semantic dependencies within each modality.

Since RWKV does not naturally provide a cross-attention mechanism for multimodal interaction, we adopt a FiLM-based conditioning strategy to perform modality-aware feature modulation. The conditioned representations from different modalities are subsequently fused into a unified joint representation space through multimodal fusion. During pretraining, the shared representation is optimized through multimodal self-supervised objectives. For downstream evaluation, the same backbone design is adapted to each task and trained from scratch using task-specific input modalities and prediction heads.

In addition, we investigate an enhanced variant of MusicRWKV by incorporating Soft ROSA into the RWKV encoders. Soft ROSA introduces differentiable neural-symbolic interactions through structured retrieval and aggregation over historical candidate representations. From a complementary perspective, Soft ROSA can be viewed as a router-free expert selection mechanism, where a subset of highly relevant candidates is selectively emphasized without requiring explicit routing networks. We compare the standard MusicRWKV and the Soft ROSA enhanced variant through a controlled pretraining ablation and analyze the retrieval behavior of Soft ROSA using layer-wise telemetry statistics.

Formally, given waveform, spectrogram, and metadata token sequences

$$X = \{X_w, X_s, X_m\}$$

MusicRWKV learns a unified representation

$$H = f_{\text{MusicRWKV}}(X_w, X_s, X_m)$$

where $H$ is optimized through multimodal self-supervised objectives. For downstream tasks, the MusicRWKV backbone architecture is reused with task-specific modality configurations and output heads, while model parameters are initialized and trained from scratch.

## 3.2 Multimodal Input Representation

Music contains heterogeneous information spanning acoustic, spectral, and semantic levels. To capture complementary musical characteristics, MusicRWKV jointly models waveform, spectrogram, and metadata modalities. These modalities are converted into token sequences and subsequently processed by modality-specific branches.

For waveform modeling, raw audio recordings are tokenized using the Jukebox [13] tokenizer, which converts continuous audio signals into discrete codec tokens while preserving fine-grained temporal structures. To provide complementary spectral information, we further extract spectrogram tokens using the MuQ Mel-RVQ tokenizer. Compared with waveform tokens, Mel-RVQ tokens emphasize timbral and harmonic characteristics that are important for music understanding tasks.

In addition to acoustic representations, we incorporate semantic metadata including genre, artist, and lyrics. These metadata tokens provide high-level contextual information that may not be directly observable from audio signals alone.

Among the three modalities, waveform and spectrogram sequences are typically much longer than metadata sequences. To improve computational efficiency, waveform and spectrogram tokens are first passed through an invertible audio compression module before representation learning, while metadata tokens are directly forwarded to the corresponding encoder. The compression module groups adjacent tokens into fixed-length segments and projects them into a latent representation, while an inverse projection module reconstructs the original token sequence during training. The resulting compressed and uncompressed token sequences serve as inputs to the MusicRWKV architecture.

## 3.3 Architecture

Given the multimodal inputs $X = \{X_w, X_s, X_m\}$, MusicRWKV learns modality-aware music representations through compression, RWKV encoding, conditioning, and multimodal fusion, as illustrated in Figure 1.

Since waveform and spectrogram token sequences are considerably longer than metadata sequences, directly processing them may introduce unnecessary computational overhead. To improve efficiency, waveform and spectrogram tokens are first passed through an invertible audio compression module, while metadata tokens are directly forwarded to the corresponding encoder. Let

$$Z_{w} = C(X_{w}) \qquad Z_{s} = C(X_{s})$$

where $C(\cdot)$ denotes the compression function. The compressed representations preserve essential musical information while reducing sequence length for subsequent representation learning.

The resulting modality-specific token sequences are then processed by independent RWKV encoders. Unlike Transformer architectures that rely on self-attention, RWKV combines Transformer-style parallel training with recurrent inference, enabling efficient long-context modeling with linear inference complexity. Such characteristics are particularly desirable for music understanding, where recordings often contain long temporal dependencies spanning rhythm, timbre, harmony, and structural information.

Let $f_{\text{rwkv}}$ denote a modality-specific RWKV encoder. The encoded representations can be written as

$$H_{w} = f_{\text{rwkv}}(Z_{w})$$
$$H_{s} = f_{\text{rwkv}}(Z_{s})$$
$$H_{m} = f_{\text{rwkv}}(X_{m})$$

Since RWKV does not naturally provide a cross-attention mechanism for multimodal interaction, we employ a FiLM-based conditioning strategy to inject modality-specific information into the learned representations. Following the FiLM formulation, feature modulation is performed through learnable affine transformations,

$$FiLM(H) = \gamma \odot H + \beta$$

where $\gamma$ and $\beta$ are conditioning parameters determined by the corresponding modality.

The conditioned representations are subsequently fused to obtain a unified music representation,

$$H_{joint} = fusion(H_w, H_s, H_m)$$

where $fusion$ denotes the multimodal fusion operation. In our implementation, the fusion operation is realized by concatenating modality-specific representations followed by a shared RWKV encoder.

## 3.4 Soft ROSA Enhanced Variant

To further investigate the effect of neural-symbolic interaction in music representation learning, we introduce an enhanced variant of MusicRWKV by incorporating the Soft ROSA mechanism into modality-specific RWKV encoders. Unlike the standard MusicRWKV framework, the enhanced variant augments representation learning through differentiable retrieval and aggregation over historical candidate representations while preserving end-to-end differentiability.

Soft ROSA operates on intermediate hidden representations. Given query, key, and value representations $Q$, $K$, and $V$, the Soft ROSA module performs structured candidate matching and aggregation to produce enriched representations

$$H' = SoftROSA(Q, K, V)$$

which are subsequently propagated through the remaining RWKV layers.

From a representation learning perspective, Soft ROSA can be interpreted as a differentiable neural-symbolic interaction mechanism that enables hidden representations to selectively retrieve information from relevant historical contexts. Such interactions provide an additional structured retrieval process beyond the recurrent dynamics of RWKV and may facilitate the modeling of high-level musical concepts and long-range dependencies.

From a complementary perspective, Soft ROSA can also be viewed as a router-free expert selection mechanism. Instead of dispatching tokens through explicit routing networks, the model progressively emphasizes a subset of highly relevant historical candidates through differentiable matching and weighting operations.

In this work, we employ Soft ROSA as an optional enhancement to MusicRWKV and investigate its influence on multimodal pretraining through a controlled ablation study.

## 3.5 Training Objective

MusicRWKV is trained through multimodal self-supervised objectives. Given waveform, spectrogram, and metadata inputs, the model learns modality-specific and joint representations through modality-aware encoding and fusion.

The primary training objective consists of modality-level cross entropy losses over waveform, spectrogram, and metadata representations,

$$L_{main} = L_{wave} + L_{spec} + L_{meta}$$

To mitigate information loss introduced by compression and modality fusion, we introduce a lightweight reconstruction regularization term. Specifically, the reconstructed waveform and spectrogram representations are encouraged to remain close to their original inputs,

$$L_{recon} = \left\|X_{w} - \hat{X}_{w}\right\|_2^2 + \left\|X_{s} - \hat{X}_{s}\right\|_2^2$$

Unlike modality-level prediction objectives, the reconstruction loss does not serve as an independent learning target. Instead, it acts as an information-preserving regularization that encourages reconstructed representations to remain close to their original embedded representations.

The final training objective is defined as

$$L = L_{main} + \lambda L_{recon}$$

where $\lambda$ controls the contribution of the reconstruction regularization term.

# 4 Experiments

## 4.1 Experimental Setup

### Pretraining Configuration

MusicRWKV was pretrained on approximately 25,000 publicly accessible full-length music recordings collected from Internet sources. To encourage stylistic diversity, the corpus was assembled with approximately balanced genre coverage whenever possible. The dataset was partitioned into training and validation subsets using an 80/20 genre-stratified split. To prevent evaluation leakage, recordings overlapping with any downstream benchmark datasets were excluded from the pretraining corpus.

Following the architecture described in Section 3, raw audio recordings were converted into discrete token sequences using the Jukebox tokenizer and the MuQ Mel-RVQ tokenizer. Waveform tokens preserve fine-grained temporal information, whereas spectrogram tokens provide complementary spectral characteristics such as timbre and harmonic structure. Metadata information, including artist, genre, and lyrics when available, was converted into textual token sequences and jointly modeled with audio representations. The maximum sequence lengths were set to 20,000 waveform tokens, 10,000 spectrogram tokens, and 500 metadata tokens. Such sequence lengths highlight the long-context nature of multimodal music modeling and motivate the use of RWKV as the backbone architecture.

The model was pretrained using the multimodal self-supervised objectives described in Section 3.4, consisting of waveform, spectrogram, and metadata prediction losses together with a lightweight reconstruction regularization term. Unless otherwise specified, both MusicRWKV and the Soft ROSA enhanced variant were trained under identical optimization settings, with the inclusion of the Soft ROSA module being the only architectural difference between the two models. Detailed optimization settings are provided in Appendix C.

Table 1. Architecture Configuration of the Pretraining MusicRWKV Model

| Component | Configuration |
|---|---|
| Hidden Dimension | 1024 |
| Waveform RWKV Layers | 6 |
| Spectrogram RWKV Layers | 6 |
| Metadata RWKV Layers | 6 |
| Fusion RWKV Layers | 4 |
| Compression Stride | 10 |
| Trainable Parameters | 883M |

The configuration in Table 1 corresponds to the multimodal pretraining model. The downstream models use smaller task-specific configurations with reduced model capacity and hidden dimensionality, while retaining the same general RWKV backbone design. The parameter count includes modality-specific embedding layers, invertible compression modules, RWKV encoders, FiLM conditioning layers, and output prediction heads. A substantial portion of the parameters originates from the modality-specific compression modules introduced for invertible sequence compression.

**Downstream Tasks**

To evaluate the applicability of the proposed RWKV-based architecture, we conduct experiments on a diverse collection of downstream Music Information Retrieval (MIR) tasks covering music tagging, genre recognition, musical attribute prediction, and singing voice analysis.

Specifically, we evaluate MusicRWKV on the following benchmark datasets:

- **MagnaTagATune (MTT)** [14] for automatic music tagging, where the objective is to predict semantic tags describing musical content.
- **GTZAN** [15] for music genre classification.
- **GiantSteps Key** [16] for musical key estimation.
- **EMO** [17] for music emotion recognition.
- **NSynth** [18] for instrument classification and pitch classification.
- **VocalSet** [19] for singing voice analysis, including singer identification and vocal technique classification.

These benchmarks cover a broad range of MIR scenarios, ranging from high-level semantic understanding to low-level acoustic attribute prediction. Consequently, they provide a broad evaluation of the applicability of the proposed architecture across heterogeneous MIR tasks.

For downstream evaluation, we reuse the general MusicRWKV backbone design but train each task-specific model from scratch without loading the pretrained MusicRWKV parameters. Compared with the pretraining model, the downstream configurations use fewer parameters and lower-dimensional hidden representations to reduce overfitting on relatively small, labeled datasets. The input modalities and output heads are adapted to the characteristics of each task. Accordingly, the downstream experiments evaluate the applicability of the MusicRWKV architecture across heterogeneous MIR tasks rather than a conventional frozen-probing or fine-tuning protocol.

Detailed input modalities, hidden dimensions, backbone depths, parameter counts, and prediction heads for individual downstream tasks are provided in Appendix B.

**Baseline Models**

We compare MusicRWKV against representative self-supervised music representation learning models spanning contrastive learning, masked prediction, multitask learning, and large-scale music foundation model paradigms. Specifically, we include CLMR and MULE as early self-supervised music representation learning baselines, MERT, MusicFM, and MuQ as representative Transformer-based music foundation models, MT2 as a recent multitask self-supervised learning framework, and AudioMAE++ [20] as a strong reconstruction-based audio foundation model. In addition, the effect of Soft ROSA is evaluated separately through a controlled pretraining ablation and layer-wise telemetry analysis.

Table 2. Representative Baseline Models and Their Learning Paradigms

| **Category** | **Models** |
|---|---|
| Contrastive SSL | CLMR, MULE |
| Masked Prediction | MERT, MusicFM |
| Token Modeling | MuQ |
| Multitask SSL | MT2 |
| Reconstruction-based SSL | AudioMAE++ |

**Evaluation Metrics**

We follow the standard evaluation protocols adopted by each benchmark. Depending on the task, performance is measured using ROC-AUC, Average Precision (AP), Refined Accuracy, Accuracy, or $R^2$. Higher values indicate better performance.

## 4.2 Main Results

Tables 3 and 4 summarize the downstream evaluation results of MusicRWKV on a diverse collection of MIR benchmarks. Overall, MusicRWKV achieves competitive performance across both general music understanding and fine-grained musical attribute prediction tasks, demonstrating the broad applicability of the proposed backbone design. Particularly strong performance is observed on music tagging average precision, musical key estimation, and vocal technique classification, indicating that task-specific variants of the proposed backbone can capture both high-level semantic information and fine-grained expressive musical characteristics.

Table 3. Performance Comparison on General MIR Benchmarks

| **Dataset** | **MTT** | | **GS** | **GTZAN** | **EMO** | |
|---|---|---|---|---|---|---|
| **Task** | tagging | | key | genre | emotion | |
| **Metrics** | ROC | AP | $Acc^{refined}$ | Acc | valence/$R^2$ | arousal/$R^2$ |
| CLMR | 89.4† | 36.1† | 14.9† | 68.6† | 45.8† | 67.8† |
| MULE | 91.4† | 40.4† | 66.7† | 73.5† | 57.7† | 70.0† |
| MERT | 90.6‡ | 37.90‡ | 64.1‡ | 77.6‡ | 56.7‡ | 76.1‡ |
| MERT2-30s | **91.9**‡ | 41.3‡ | 66.9‡ | **91.7**‡ | 63.2‡ | **80.0**‡ |
| MERT2-FS | 91.7‡ | 41.2‡ | 67.1‡ | 90.7‡ | **63.5**‡ | 78.1‡ |
| MusicFM | 90.9‡ | 38.3‡ | 63.0‡ | 84.1‡ | 57.2‡ | 74.4‡ |
| MuQ | 90.5‡ | 38.5‡ | 63.2‡ | 83.8‡ | 58.3‡ | 76.4‡ |
| MT2 | 88.4† | 38.8† | - | - | - | - |
| AudioMAE++ | 91.2‡ | 39.5‡ | 61.7‡ | 80.3‡ | 59.0‡ | 75.7‡ |
| MusicRWKV | 88.6 | **46.1** | **70.3** | 86.6 | 62.1 | 75.1 |

Table 4. Performance on Fine-Grained Musical Attribute and Singing Voice Tasks

| **Dataset** | **Nsynth** | | **VocalSet** | |
|---|---|---|---|---|
| **Task** | instrument | pitch | technique | singer |
| **Metrics** | Acc | Acc | Acc | Acc |
| CLMR | 67.9† | 47.0† | 58.1† | 49.9† |
| MULE | **74.0**† | 89.2† | 75.5† | **87.5**† |
| MERT | 72.6† | **94.4**† | 76.9† | 87.1† |
| MusicRWKV | 73.3 | 87.4 | **89.1** | 83.0 |

† Results reported in the original publication.

‡ Results obtained from the public MARBLE benchmark.

MusicRWKV results are obtained by training task-specific models from scratch using the proposed backbone architecture. Baseline results are taken from original publications or the public MARBLE benchmark and may follow pretrained probing protocols. Therefore, these results should be interpreted as reference comparisons rather than strictly controlled evaluations under an identical training paradigm.

An important practical observation is that directly transferring the large pretrained MusicRWKV model to downstream benchmarks frequently resulted in substantial overfitting. This effect was particularly evident for datasets with limited labeled samples and appears to arise from the mismatch between the capacity and high-dimensional representation space of the pretrained model and the scale of downstream supervision. In preliminary experiments, reducing both model capacity and representation dimensionality resulted in more stable optimization and improved generalization. We therefore adopted smaller task-specific configurations and trained all downstream models from scratch.

Another notable finding is that downstream performance depends strongly on task-specific modality and representation choices. The evaluated tasks emphasize different musical factors, including harmonic structure, timbre, pitch, vocal expression, and performer identity. MusicRWKV performs particularly well on key estimation and vocal technique classification, whereas pitch classification and singer identification remain more challenging. These results suggest that MIR tasks may benefit from task-aware modality configurations and representation designs rather than a single fixed input pipeline.

## 4.3 Pretraining Ablation Study

To evaluate the impact of Soft ROSA on music representation learning, we conduct a pretraining ablation study by comparing the standard MusicRWKV framework with its Soft ROSA enhanced variant. Both models are trained under identical settings, with the only difference being the inclusion of the Soft ROSA module within the RWKV encoder.

Table 5. Effect of Soft ROSA on Pretraining Performance

| Model | Params (M) | Peak Memory (GiB) | Time / Sample | Validation Accuracy (%) |
|---|---|---|---|---|
| MusicRWKV | 883.00 | 64.90 | 3.32 s | 93.30 |
| MusicRWKV + Soft ROSA | 952.66 | 69.80 | 9.49 s | 94.82 |

Both models were trained using identical data splits, initialization seeds, and optimization settings. The reported validation accuracies correspond to the selected checkpoint with the lowest validation loss for each model.

As shown in Table 5, incorporating Soft ROSA improves validation accuracy from 93.3% to 94.8% under identical training settings. Since both models were trained using the same data split, random seed, and optimization configuration, the observed improvement is consistent with the inclusion of the Soft ROSA retrieval mechanism.

To better understand the source of this improvement, we further analyze the behavior of Soft ROSA through layer-wise telemetry statistics in the next section.

## 4.4 Empirical Analysis of Soft ROSA

### Layer-wise Telemetry

Telemetry statistics are computed from the normalized routing distributions produced by Soft ROSA and averaged across all batch elements, heads, and token positions. Top Probability denotes the maximum routing probability, Entropy corresponds to the normalized Shannon entropy of the routing distribution, and Candidate Mass denotes the total probability assigned to retrieved candidates (equivalently, one minus the sink probability). Table 6 summarizes the resulting layer-wise statistics.

$$p_{top} = \max_i p_i$$

$$M_c = \sum_i p_i = 1 - p_{sink}$$

$$H_{norm} = \frac{-\sum_i p_i \log p_i}{\log K}$$

where `$p_i$ denotes the normalized routing probability, $p_{sink}$ denotes the sink-route probability, and $K$ is the number of available routes. The entropy metric is normalized to the range $[0,1]$.

Table 6. Layer-wise Telemetry Statistics of Soft ROSA

| Layer | Top Probability | Entropy | Candidate Mass |
|---|---|---|---|
| 0 | 0.31 | 0.73 | 0.86 |
| 1 | 0.38 | 0.65 | 0.82 |
| 2 | 0.47 | 0.56 | 0.79 |
| 3 | 0.58 | 0.45 | 0.73 |
| 4 | 0.71 | 0.31 | 0.66 |
| 5 | 0.84 | 0.18 | 0.58 |

As layer depth increases, the top probability consistently rises from 0.31 to 0.84, while routing entropy decreases from 0.73 to 0.18. In addition, candidate mass gradually decreases from 0.86 to 0.58. In our experiments, these trends indicate that retrieval distributions become progressively more concentrated in deeper layers. Rather than distributing attention broadly across many candidates, later layers increasingly focus on a smaller subset of highly relevant historical candidates.

**Retrieval Behavior Discussion**

The telemetry analysis suggests that Soft ROSA exhibits a hierarchical retrieval behavior across network depth. Early layers maintain relatively distributed routing patterns, allowing representations to aggregate information from multiple historical candidates. As representations become progressively refined, deeper layers increasingly emphasize a smaller subset of highly relevant candidates.

This behavior is consistent with the interpretation of Soft ROSA as a differentiable retrieval mechanism operating over historical candidate representations. Rather than relying on explicit routing networks, Soft ROSA gradually concentrates probability mass onto selected candidates through differentiable matching and aggregation. Such behavior aligns with the router-free expert-selection perspective introduced in Section 3 and may contribute to the improved pretraining performance observed in the ablation study.

Additional qualitative visualization of the learned representation space using UMAP is provided in Appendix A.

# 5 Conclusion

In this work, we presented MusicRWKV, a multimodal RWKV framework for music representation learning. Unlike existing music foundation models that primarily rely on Transformer architectures, MusicRWKV adopts RWKV as the representation backbone and jointly models waveform, spectrogram, and metadata information within a unified multimodal framework. To facilitate multimodal interaction without introducing expensive cross-attention operations, we employed FiLM-based conditioning and investigated the integration of Soft ROSA as an optional differentiable retrieval mechanism.

Experimental results across a diverse collection of MIR benchmarks demonstrate that the proposed RWKV-based backbone can be effectively adapted to heterogeneous music understanding tasks. These downstream models share the same general backbone design but are trained from scratch using task-specific input modalities and output heads. In the pretraining study, incorporating Soft ROSA improves validation performance, while layer-wise telemetry analysis reveals a progressive transition from distributed retrieval to selective retrieval.

Several limitations remain. First, the retrieval mechanism introduced by Soft ROSA incurs additional computational and memory overhead, which may limit scalability for long-context music modeling. Second, the present pretraining study evaluates a single large-scale MusicRWKV configuration and does not systematically explore scaling behavior across different pretraining model sizes. Understanding how representation quality varies with model scale therefore remains an important direction for future work. Finally, our results suggest that representation suitability may vary across downstream tasks, indicating that a unified multimodal representation may not be optimal for all MIR scenarios.

Future work will investigate more efficient retrieval mechanisms for Soft ROSA, larger-scale pretraining strategies, and adaptive modality selection methods tailored to different music understanding tasks. We believe that these directions will further improve the scalability and cross-task applicability of RWKV-based music models.

# Appendix A

## Representation Space Visualization

To provide additional qualitative insight into the learned representation space, we visualize MusicRWKV embeddings using UMAP. Figure A1 shows UMAP visualizations of VocalSet representations colored by singer identity and vocal technique.

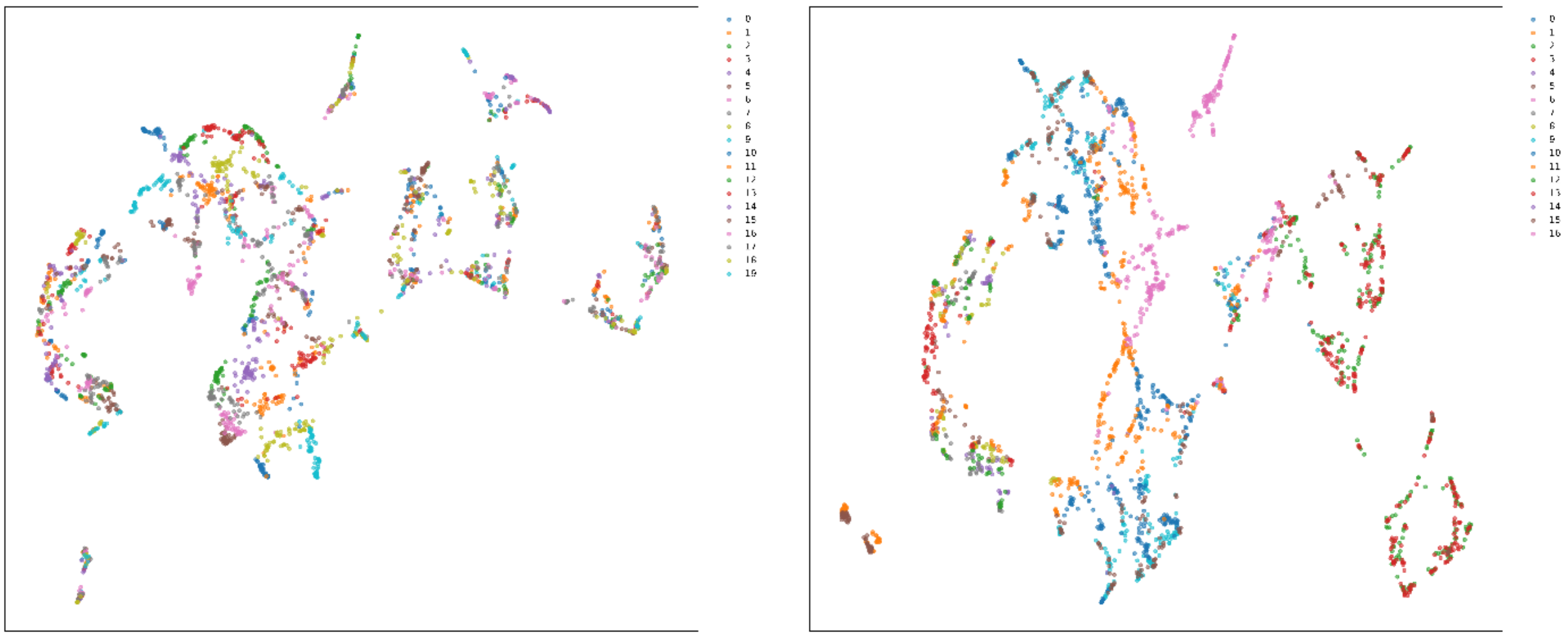

(a) colored by singer identity (b) colored by vocal technique

Figure A1. UMAP visualization of VocalSet representations

The visualizations suggest slightly stronger local organization with respect to vocal techniques than singer identities, which is broadly consistent with the downstream classification results reported in Table 4. While the overall clustering structure remains diffuse, the embeddings reveal meaningful organization related to vocal expression.

# Appendix B

## Downstream Task Configurations

All downstream models reuse the general MusicRWKV backbone design but are trained from scratch without loading pretrained MusicRWKV parameters. Task-specific configurations differ in input modality selection, backbone depth, parameter count, and prediction heads according to the characteristics of each benchmark.

Table B1. Task-Specific Downstream Model Configurations

| **Dataset** | **Task** | **Input Modalities** | **Hidden Dim** | **RWKV Layers** | **Params (M)** | **Output Head** | **Initialization** |
|---|---|---|---|---|---|---|---|
| MTT | Tagging | Waveform + Spectrogram | 512 | 6 | 21.20 | Multi-label Classification | From Scratch |
| GiantSteps | Key | Harmonic Mel Spectrogram | 512 | 2 | 19.70 | Classification | From Scratch |
| GTZAN | Genre | Waveform + Spectrogram | 512 | 6 | 21.33 | Classification | From Scratch |
| EMO | Emotion | Waveform + Spectrogram | 512 | 6 | 21.36 | Regression | From Scratch |
| NSynth | Instrument | Spectrogram | 512 | 6 | 21.54 | Classification | From Scratch |
| NSynth | Pitch | Harmonic CQT | 512 | 6 | 21.54 | Classification | From Scratch |
| VocalSet | Technique | CREPE Pitch Embeddings | 512 | 6 | 20.35 | Classification | From Scratch |
| VocalSet | Singer | WavLM Speaker Embeddings | 512 | 6 | 20.35 | Classification | From Scratch |

# Appendix C

## Pretraining Optimization Details

Table C1. Pretraining Dataset and Optimization Settings

| Item | Value |
|---|---|
| Dataset Size | ~25,000 recordings |
| Train / Validation Split | 80 / 20 |
| Split Strategy | Genre-stratified |
| Leakage Prevention | Downstream benchmark recordings excluded |
| Max Waveform Length | 20,000 |
| Max Spectrogram Length | 10,000 |
| Max Metadata Length | 500 |
| Optimizer | Adam |
| Initial Learning Rate | 1e-3 |
| Batch Size | 4 |
| Epochs | 400 |
| Scheduler | ReduceLROnPlateau |
| Scheduler Factor | 0.5 |
| Scheduler Patience | 3 |
| Minimum Learning Rate | 1e-6 |
| Checkpoint Selection | Lowest validation loss |